\documentclass[a4paper]{jpconf}

\newcommand{\mq}{\mathbf{q}}
\newcommand{\mx}{\mathbf{x}}
\newcommand{\mf}{\mathbf{f}}
\newcommand{\mg}{\mathbf{g}}
\newcommand{\mh}{\mathbf{h}}
\newcommand{\mmu}{\mathbf{u}}
\newcommand{\mmr}{\mathbf{r}}

\newcommand{\mW}{\boldsymbol{W}}

\newcommand{\p}[1]{(\ref{#1})}

\newcommand{\bD}{{\overline D}}

\newcommand{\bQ}{{\overline Q}{}}

\newcommand{\bC}{{\overline C}}

\newcommand{\bpsi}{{\bar\psi}{}}

\newcommand{\be}{\begin{equation}}
\newcommand{\ee}{\end{equation}}
\newcommand{\bea}{\begin{eqnarray}}
\newcommand{\eea}{\end{eqnarray}}

\newcommand{\ba}{\begin{array}} \newcommand{\ea}{\end{array}}

\def\im{{\rm i}}

\newcommand{\nn}{\nonumber}

\usepackage{amscd,amsmath,amssymb}

\begin{document}
\title{The curved WDVV equations and superfields}
\author{N Kozyrev}
\address{Bogoliubov  Laboratory for Theoretical Physics, JINR,
141980 Dubna, Russia}
\ead{nkozyrev@theor.jinr.ru}

\begin{abstract}\noindent
We reproduce the ${\cal N}=4$ supersymmetric mechanics on curved spaces, constructed earlier within the Hamiltonian formalism, using the superfield methods. We show that for any such mechanics, given by the metric and the third order Codazzi tensor, it is possible to construct a suitable modification of irreducibility conditions of linear ${\cal N}=4$ multiplets and obtain the superfield Lagrangian by solving a simple differential equation. Also, we prove that the constructed irreducibility conditions are consistent if and only if the metric and Codazzi tensor satisfy the modification of the WDVV equations, which are the conditions of existence of ${\cal N}=4$ supersymmetry.

\end{abstract}

\setcounter{page}{1}
\setcounter{equation}{0}

\section*{Introduction}
In a few recent papers we studied the ${\cal N}=4$ supersymmetric mechanics of a few linear multiplets with curved target spaces within the Hamiltonian formalism \cite{cWDVV1}, \cite{cWDVV2}. In such systems a pair of a physical bosonic coordinate and its conjugate momenta $\big(x^i, p_i\big)$ is accompanied by four fermionic variables $\big( \psi^{ia}, \bpsi^i_a = \big(\psi^{ia} \big){}^\dagger  \big)$, $i=1,\ldots,N$, $a=1,2$. As the Dirac brackets between these variables, without loss of generality, can be chosen as
\bea\label{xppsibracs}
&\big\{ x^i , p_j \big\}= \delta^i_j, \;\; \big\{ \psi^{ia}, \bpsi^j_b  \big\} = \im g^{ij}\delta_b^a,& \nn \\ &\big\{ p_i, \psi^{ja}  \big\} =\Gamma^j_{ik} \psi^{ka}, \;\; \big\{ p_i, \bpsi^j_a  \big\} =\Gamma^j_{ik} \bpsi^{k}_a, \;\; \big\{  p_i, p_j \big\} = -2\im R_{ijkl}\psi^{ck}\bpsi^l_c,&
\eea
most general $U(1)$ covariant supercharges, which involve terms up to cubic ones in the fermions,
\be\label{Qans}
Q^a = p_i \psi^{ai} + \im F_{(ij)k}\psi^{ci}\psi_c^j \bpsi^{ak} +\im G_{[ij]k}\psi^{ai}\psi^{cj}\bpsi_c^k, \;\; {\overline Q_a}=\big(Q^a\big)^\dagger,
\ee
form the ${\cal N}=4$, $d=1$ super-Poincar\'{e} algebra
\be\label{N4d1sP}
\big\{ Q^a, \bQ_b  \big\} =2\im H \delta_b^a, \;\; \big\{ Q^a, Q^b \big\} = \big\{ \bQ_a, \bQ_b\big\}=0
\ee
if $G_{ijk}=0$ and the following equations hold:
\be\label{cWDVV}
F_{ijk}=F_{(ijk)}, \;\; \nabla_i F_{jkl} - \nabla_j F_{ikl}=0 \;\; \mbox{and} \;\;F_{ikp}g^{pq}F_{jlq} - F_{ilp}g^{pq}F_{jkq}+R_{ijkl}=0.
\ee
Here, $\Gamma^k_{ij}$ and $R_{ijkl} = g_{im}R^m_{jkl}$ are the Levi-Civita connection and the Riemann tensor of the metric $g_{ij}$, respectively,
\be\label{GammaR}
\Gamma^k_{ij} = \frac{1}{2}g^{kl}\big( \partial_i g_{jl}+\partial_j g_{il} - \partial_l g_{ij}  \big), \;\; R^i_{jkl} = \partial_k \Gamma^i_{jl} - \partial_l \Gamma^i_{jk} + \Gamma^i_{kp}\Gamma^p_{jl} - \Gamma^i_{lp}\Gamma^p_{jk}.
\ee
In the specific case of $g_{ij} =\delta_{ij}$, the first two equations imply $F_{ijk}=\partial_i \partial_j \partial_k F$ and  the last one reduces to the well-known WDVV equation \cite{WittenWDVV},\cite{DVVWDVV}. The WDVV equation in the context of the ${\cal N}=4$ supersymmetric mechanics first appeared in \cite{Wyllard} and was noted in \cite{BGL} in the case of flat target spaces. In \cite{cWDVV1}, the system \p{cWDVV} was called ``curved WDVV equations''. Note that the first two equations \p{cWDVV} qualify $F_{ijk}$ as the so called third order Codazzi tensor \cite{Codazzi}.

If equations \p{cWDVV} hold, the Hamiltonian can be written as
\be\label{Ham}
H = g^{ij}p_i p_j - 4 \big[ \nabla_i F_{jkm} + R_{ijkm} \big]\psi{}^{ai}\bpsi{}_a^j \, \psi{}^{bk}\bpsi{}_b^m,
\ee
with $g_{ij}$ clearly appearing as the metric of the target space. A number of systems with different metrics were constructed, such as the isotropic metric $g_{ij} = f^{-2}\big(\sum_k(x^k)^2 \big)\delta_{ij}$ \cite{cWDVV1}.  Extensions of this system, which involve two types of potentials (with and without the harmonic variables) \cite{cWDVV2} and $SU(2|1)$ supersymmetry \cite{cWDVVSU21} were also studied.

Let us note, however, that the systems of $N=4$ linear multiplets on curved target spaces were already studied in \cite{Pash1}, \cite{Pash2} using the superfield method. In \cite{Pash1}, the conclusion was reached that the admissible metrics are the second derivatives of the prepotential $g_{ij} = \partial_i \partial_j G$ (which in \cite{Pash1} was called the ``real K\"{a}hler metric''), while $F_{ijk} = \frac{1}{2}\partial_i\partial_j \partial_k G$ (these $g_{ij}$ and $F_{ijk}$ provide one set of particular solutions of equations \p{cWDVV}). The choice of potentials was also limited. At the same time, the curved WDVV equations are meant to be solved w.r.t. $F_{ijk}$ and thus provide no obvious constraint on the metric. Moreover, they were solved on the spaces the metrics of which do not satisfy  $g_{ij} = \partial_i \partial_j G$. However, the definition of real K\"{a}hler metrics is not coordinate independent, and it is difficult to show for any given metric whether it can be presented in this particular form in some system of coordinates or not. These results naturally raise the question whether the two results are really contradicting and, if so, what superfield techniques are required to reproduce the systems, which are possible to construct within the Hamiltonian formalism.

In this paper, we show that for all mechanics given by the solutions of the curved WDVV equations, it is possible to provide a superfield Lagrangian, together with semitrivial deformation of the linear multiplet constraints, which reproduce these mechanics exactly. Moreover, as any of the superfield mechanics provide a solution to the curved WDVV equations, this establishes the two-way equivalence between these formalisms. As a side result of this short proof, we propose the covariant definition of the real K\"{a}hler spaces. To simplify our construction, we consider only standard potentials and do not make an attempt to introduce the potentials induced by the harmonic variables \cite{FILharms}, \cite{KLD21alpha}.

\section{New irreducibility conditions}
The standard $N=4$ linear multiplets are defined by the relations
\be\label{linm1}
D^a D_a \mx^i = 0, \;\; \bD_a \bD^a \mx^i =0, \;\; \big[ D^a, \bD_a  \big]\mx^i =8n^i.
\ee
(It is actually optional to add constants to the right-hand sides of all three relations without making the irreducibility conditions inconsistent and without putting them on-shell). However, if one uses these conditions to find the component form of the most general possible bosonic action, the resulting metric would inevitably be a second derivative \cite{Pash1}:
\bea\label{linmact1}
S&=& - \int dt d^4 \theta G(\mx^i)=-\frac{1}{16} \int dt D^a D_a \bD_b \bD^b G(\mx^i) = \nn \\&=& \int dt \partial_i \partial_j G(x^k)\Big( \frac{1}{4} {\dot x}{}^i {\dot x}{}^j - n^i n^j   \Big) + \mbox{fermions},
\eea
where $x^i = \mx^i |_{\theta\rightarrow 0} $ and the auxiliary field was removed by its algebraic equation of motion. The expression for the metric can obviously be changed by passing to another coordinate system.  The same change of coordinates $x^i = x^i(x^{\prime j})$ can also be applied not just to the first components but also to the superfields $\mx^i = \mx^i(\mx^{\prime j})$. In this case, the irreducibility conditions would be deformed:
\be\label{linm2}
\frac{\partial \mx^i}{\partial \mx^{\prime k}} D^a D_a \mx^{\prime k} + \frac{\partial^2 \mx^i}{\partial \mx^{\prime j} \partial \mx^{\prime k}} D^a \mx^{\prime j} D_a \mx^{\prime k}=0, \;\; \mx^i = \mx^i(\mx^{\prime j}).
\ee

One may, therefore, follow \cite{KLP} and propose more general irreducibility conditions
\bea\label{genxconstr}
&D^a D_a \mx^i +\mf^{i}_{jk}D^a \mx^j D_a \mx^k =0,& \nn \\
&\bD_a \bD^a \mx^i +\mf^{i}_{jk}\bD_a \mx^j \bD^a \mx^k =0,& \\
&\big[ D^a, \bD_a \big]\mx^i + 2 \mh^i_{jk} D^a \mx^j \bD_a \mx^k +8 \mW^i=0.&\nn
\eea
If we assume that these relations are covariant with respect to the transformations $\mx^{ i} = \mx^{i}(\mx^{\prime j})$, then $\mW^i$ should transform as a vector, while $\mf^i_{jk}$ and $\mh^i_{jk}$ should exhibit the properties of the connections. Indeed, as
\bea\label{genxcontsrvar}
D^a D_a \mx^i = \frac{\partial \mx^i}{\partial {\mx^\prime}^{i^\prime}}D^a D_a  {\mx^\prime}^{i^\prime} + \frac{\partial^2 \mx^i}{ \partial{\mx^\prime}^{j^\prime} \partial {\mx^\prime}^{k^\prime}}D^a  {\mx^\prime}^{j^\prime}  D_a {\mx^\prime}^{k^\prime}, \nn \\
\mf^i_{jk} D^a \mx^j D_a \mx^k =\mf^i_{jk}   \frac{\partial \mx^j}{\partial {\mx^\prime}^{j^\prime}}  \frac{\partial \mx^k}{\partial {\mx^\prime}^{k^\prime}}  D^a {\mx^\prime}^{j^\prime} D_a{\mx^\prime}^{k^\prime},
\eea
the first condition is covariant if
\bea\label{genxcov}
D^a D_a \mx^i +\mf^{i}_{jk}D^a \mx^j D_a \mx^k = \frac{\partial \mx^i}{\partial {\mx^\prime}^{i^\prime}} \big( D^a D_a {\mx^\prime}^{i^\prime} +{\mf^\prime}^{i^\prime}_{j^\prime k^\prime } D^a {\mx^\prime}^{j^\prime} D_a{\mx^\prime}^{k^\prime} \big) \Rightarrow \nn \\
{\mf^\prime}^{i^\prime}_{j^\prime k^\prime } = \frac{\partial {\mx^\prime}^{i^\prime}}{\partial \mx^i} \frac{\partial \mx^j}{\partial {\mx^\prime}^{j^\prime}}  \frac{\partial \mx^k}{\partial {\mx^\prime}^{k^\prime}} \mf^i_{jk} + \frac{\partial {\mx^\prime}^{i^\prime}}{\partial \mx^i} \frac{\partial^2 \mx^i}{ \partial{\mx^\prime}^{j^\prime} \partial {\mx^\prime}^{k^\prime}}.
\eea
As
\bea\label{genxcontsrvar2}
\big[ D^a, \bD_a \big]\mx^i =\frac{\partial \mx^i}{\partial {\mx^\prime}^{i^\prime}}\big[ D^a, \bD_a \big]{\mx^\prime}^{i^\prime} + 2 \frac{\partial^2 \mx^i}{ \partial{\mx^\prime}^{j^\prime} \partial {\mx^\prime}^{k^\prime}} D^a  {\mx^\prime}^{j^\prime}  \bD_a {\mx^\prime}^{k^\prime},
\eea
the third condition is also covariant if
\be\label{hvar}
{\mh^\prime}^{i^\prime}_{j^\prime k^\prime } = \frac{\partial {\mx^\prime}^{i^\prime}}{\partial \mx^i} \frac{\partial \mx^j}{\partial {\mx^\prime}^{j^\prime}}  \frac{\partial \mx^k}{\partial {\mx^\prime}^{k^\prime}} \mh^i_{jk} + \frac{\partial {\mx^\prime}^{i^\prime}}{\partial \mx^i} \frac{\partial^2 \mx^i}{ \partial{\mx^\prime}^{j^\prime} \partial {\mx^\prime}^{k^\prime}}, \;\; \mW^i  =\frac{\partial \mx^i}{\partial \mx^{\prime i^\prime}} \mW^{\prime i^\prime}.
\ee

It is worth noting, however, that the objects $\mf^i_{jk}$, $\mh^i_{jk}$, $\mW^i$ that enter the irreducibility conditions are themselves constrained. For example, one may act by $D^a$ on the first of the conditions and use the property of the covariant derivatives $D^a D^b D_b =0$:
\be\label{constrcons1}
\big( \partial_m \mf^i_{jk} + \mf^i_{mp}\mf^p_{jk}  \big)D^a \mx^m\, D^b \mx^j \, D_b \mx^k =0\; \Rightarrow\; \partial_m \mf^i_{jk}-\partial_j \mf^{i}_{km} + \mf^i_{mp}\mf^p_{jk} -  \mf^i_{jp}\mf^p_{mk} = R^i{}_{kmj}[\mf]=0.
\ee
The connection $\mf^i_{jk}$ is therefore flat. Conditions on $\mh^i_{jk}$ and $\mW^i$ can also be obtained. Let us act by $D_a$ on the third constraint \p{genxconstr}. At first, we obtain
\bea\label{3cnstrder1}
D_a \big[ D^b,\bD_b   \big]\mx^i +2 \partial_m \mh^i_{jk}\, D_a \mx^m \, D^b \mx^j \, \bD_b \mx^k-\nn \\ - \mh^i_{jk} D^c D_c \mx^j \,\bD_a \mx^k - 2 \mh^i_{jk} D^b \mx^j \, D_a \bD_b \mx^k +8 \partial_j \mW^i D_a \mx^j =0.
\eea
To simplify this relation, we note that
\bea\label{smpl1}
&&D_a \big[ D^b,\bD_b   \big]\mx^i =-\bD_a D^b D_b \mx^i, \\ &&\mh^i_{jk}\, D^b \mx^j \, D_a \bD_b \mx^k =-\mh^i_{jk}\, D^b \mx^j\, \bD_a D_b \mx^k + \frac{1}{2} \mh^i_{jk}\,D_a \mx^j\, \big[ D^b, \bD_b \big]\mx^k.\nn
\eea
Using this together with basic constraints \p{genxconstr}, we obtain
\bea\label{3constrder2}
2 \big( \partial_m \mh^i_{jk} - \partial_k \mf^i_{jm}  + \mh^i_{pm} \mh^p_{jk} - \mh^i_{pk} \mf^p_{jm}  \big) D_a \mx^m\, D^b \mx^j \, \bD_b \mx^k + \nn \\
+ 2 \big( \mh^i_{jk} -\mf^i_{jk}  \big)D^b \mx^j \, \bD_a D_b \mx^k + 8\big( \partial_i \mW^j + \mh^i_{jk} \mW^k  \big)D_a \mx^j=0.
\eea
This formula immediately implies that $\mh^i_{jk} =\mf^i_{jk}$. Then the first line vanishes due to \p{constrcons1}, and finally we get a constraint on the vector $\mW^i$
\be\label{genxconstrW}
\partial_i \mW^j + \mf^i_{jk} \mW^k=0.
\ee
One may note that the proposed modification of the constraints is trivial as, if $\mf^i_{jk}$ is a connection with zero curvature, it can be put zero by some choice of basic fields $\mx^i =\mx^i(\mx^{\prime j})$. However, these modified constraints appear to be useful later.

\section{The modified metric}
If one repeats the calculation of the action \p{linmact1} with the new defining constraints \p{genxconstr}, the result would be just
\be\label{linmact2}
S= - \int dt d^4 \theta G(\mx^i)= \int dt g_{ij} \Big( \frac{1}{4} {\dot x}{}^i {\dot x}{}^j - W^i W^j   \Big) + \mbox{fermions}, \;\; g_{ij} = \partial_i \partial_j G - f^k_{ij}\partial_k G.
\ee
It can be shown that it is just the definition of the metric that can be used to establish the relation between the superfield formalism and the curved WDVV equations. Let us use the idea of \cite{KLP} and calculate the derivative of this metric:
\be\label{dgdef}
\partial_m g_{ij} = \partial_{i}\partial_j \partial_m G - g_{km}f^k_{ij} - \partial_l G \big( \partial_m f^l_{ij} + f^l_{km}f^k_{ij}   \big),
\ee
where we used the definition of the metric once again to express $\partial_i \partial_j G$ in terms of $g_{ij}$.
Now antisymmetrizing this relation with respect to $j,m$ and using the zero-curvature condition \p{constrcons1}, one may obtain
\be\label{fijm}
f^k_{im}g_{jk}-f^k_{ij}g_{km} = \partial_m g_{ij} - \partial_i g_{jm}.
\ee
If one now defines the object $F_{ijk}$ by the relation $f^k_{ij} = \Gamma^k_{ij} + g^{km}F_{ijm}$, where $F_{ijk}$ is symmetric with respect to the first pair of indices, then equation \p{fijm} would imply just
\be\label{Fijm}
-F_{ijm}+F_{jmi}=0,
\ee
making $F_{ijk}$ completely symmetric. After substitution of $f^k_{ij} =F^k_{ij}+\Gamma^k_{ij}$ into the flatness condition \p{constrcons1} one may find the equation
\be\label{Feq}
\nabla_k F^i_{jm} - \nabla_m F^i_{jk} +R^i{}_{jkm} +F^i_{pk} F^p_{jm} - F^i_{pm} F^p_{jk}=0,
\ee
that, after lowering the index $i$, splits into two, symmetric and antisymmetric in $i,j$. They resemble \p{cWDVV}:
\be
\nabla_k F_{ijm} - \nabla_m F_{ijk}=0, \;\;  R_{ijkm} +F_{ikp} F^p_{jm} - F_{imp} F^p_{jk}=0.
\ee
The equation for potential also follows after the substitution of $f^k_{ij} =F^k_{ij}+\Gamma^k_{ij}$ in \p{genxconstrW} and lowering one index:
\be\label{eqpot}
\nabla_i W_j + F_{ijk}g^{km}W_m =0.
\ee
It additionally implies, as $F_{ijk}$ is symmetric and $\nabla_i W_j - \nabla_j W_i=0$, that $W_i =\partial_i W$. The potential, therefore, is related to the scalar function in the usual way. As a result, the Hamiltonian and the supercharges, derivable from the complete component action, should have the same form as ones obtained with the use of the purely Hamiltonian formalism.

Let us note that this reasoning can be reversed. If one manages to find some third order symmetric tensor $F_{ijk}$, which  satisfies the curved WDVV equations \p{cWDVV} on space with some prescribed metric $g_{ij}$, then the object $f^k_{ij} = \Gamma^k_{ij} + g^{km}F_{ijm}$ constructed from them will have zero curvature
\be\label{Feq2}
R^i{}_{jkm}[f] =\nabla_k F^i_{jm} - \nabla_m F^i_{jk} +R^i{}_{jkm} +F^i_{pk} F^p_{jm} - F^i_{pm} F^p_{jk}=0.
\ee
Then one may extend it to the superfield $\mf^i_{jk}$ and use it to define the constraints on the superfields $\mx^i$ and, by solving the equation $\mg_{ij} = \partial_i \partial_j G - \mf^k_{ij}\partial_k G$,  find the superfield Lagrangian that reproduces the action for this system.

\section{The complete action and the Hamiltonian}
One may evaluate the integral \p{linmact2} taking into account all the fermions. It can be expressed in terms of components
\bea\label{comps}
x^i = \mx^i |_{\theta\rightarrow 0}, \; \psi^{ai} = - \frac{\im}{2} D^a \mx^i|_{\theta\rightarrow 0}, \; \bpsi_a^i = - \frac{\im}{2}\bD_{a}\mx^i |_{\theta\rightarrow 0}, \nn \\
A^{i(ab)} = \big[   D^{(a},\bD^{b)}\big] \mx^i|_{\theta\rightarrow 0}+2 \big( \Gamma^i_{pq} - F^i_{pq}  \big)D^{(a}\mx^p \, \bD^{b)}\mx^q |_{\theta\rightarrow 0}.
\eea
Using these components, the action can be written as $S = \int dt {\cal L}$, where
\bea\label{action}
{\cal L} &=& \frac{1}{4} g_{ij}\dot{x}{}^i \dot{x}{}^j  - \im g_{ij} \big( \dot{\psi}{}^{ia} \bpsi{}^{j}_a - \psi{}^{ia}\dot{\bpsi}{}^j_a  \big) - g_{ij} W^i W^j - \frac{1}{32} g_{ij}A^{i(ab)}A^j_{(ab)} +\nn \\
&&+ \im \big( g_{ip}\Gamma^p_{jk} - g_{jp}\Gamma^p_{ik}  \big)\dot{x}{}^k \psi{}^{ai}\bpsi{}_a^j-4 \nabla_i W_j\, \psi{}^{ai}\bpsi{}_a^j + 4 \big[ \nabla_i F_{jkm} + R_{ijkm} \big]\psi{}^{ai}\bpsi{}_a^j \, \psi{}^{bk}\bpsi{}_b^m.
\eea
With the definition of auxiliary field \p{comps}, its equation of motion is just $A^{i(ab)}=0$.

The Hamiltonian for the Lagrangian \p{action} can also be found. Let us define the momenta and the  canonical Poisson brackets
\be\label{momentabr}
{\tilde p}_i  = \frac{\partial {\cal L}}{\partial x^i}, \; \big\{ {\tilde p}_i, x^j  \big\}_P = \delta_i^j, \; \big(p_\psi\big)_{ai} =  \frac{\partial {\cal L}}{\partial \dot{\psi}{}^{ia}} = - \im g_{ij}\bpsi{}^j_a, \; \big(p_\bpsi\big)_{i}^a =  \frac{\partial {\cal L}}{\partial \dot{\bpsi}{}^{a}_i} = - \im g_{ij}\psi{}^{ja}.
\ee
As the fermionic momenta are expressed in terms of the fermions themselves, the system is constrained.
If the fermions obey the standard brackets $\big\{\psi{}^{ia}, \big(p_\psi\big)_{jb} \big\}_P = \delta^a_b \delta^i_j$, $\big\{\bpsi{}^{i}_a, \big(p_\bpsi\big)_{j}^b \big\}_P = \delta^b_a \delta^i_j$, the bracket of the constraints remains nontrivial and could not be treated as the third constraint:
\be\label{constr}
C_{ia} = \big(p_\psi\big)_{ia} + \im g_{ij}\bpsi{}^j_a \approx 0, \; \bC_i^a = \big(p_\bpsi\big)_{i}^a + \im g_{ij}\psi{}^{ja}\approx 0, \; \mbox {and} \; \big\{ C_{ia},\bC_j^b  \big\}_P = 2\im g_{ij}\delta^b_a.
\ee
The constraints, therefore, are those of the second class and the new Dirac brackets respecting them have to be defined. They can be calculated with the help of the formula
\be\label{drcbrdef}
\big\{ f,g  \big\} = \big\{ f,g \big\}_P - \big\{ f,C_A \big\}_P \big( O^{-1} \big)^{AB} \big\{ C_B, g \big\}_P, \; O_{AB} = \big\{ C_A, C_B  \big\}_P.
\ee
The brackets between the basic variables are
\bea\label{tildabracs}
\big\{ \psi{}^{ia}, \bpsi{}^j_b  \big\} = \frac{\im}{2} g^{ij}\delta_b^a, \; \big\{ {\tilde p}_i, \psi{}^{ja} \big\} = - \frac{1}{2} g^{jm}\partial_i g_{mp} \psi{}^{pa}, \; \big\{ {\tilde p}_i, \bpsi{}^{j}_a \big\} = - \frac{1}{2} g^{jm}\partial_i g_{mp} \bpsi{}^{p}_a, \nn \\ \big\{ {\tilde p}_i, {\tilde p}_j \big\} = \frac{\im}{2} g^{km}\big( \partial_i g_{mp}\; \partial_j g_{kq} - \partial_j g_{mp}\; \partial_i g_{kq}\big)\psi{}^{ap}\bpsi{}_a^q.
\eea
They are not of expected geometric form. However, one may define new momenta that produce right brackets
\be\label{rightp}
p_i =-{\tilde p}_i + \im \big( g_{pn} \Gamma^n_{iq} - g_{qn}\Gamma^n_{ip}  \big)\psi{}^{cp}\,\bpsi{}^q_c.
\ee
Then the previously postulated brackets can be recovered:
\bea\label{finalbracs}
&\big\{ x^i, p_j  \big\} =\delta_j^i, \; \big\{ \psi{}^{ia}, \bpsi{}^j_b  \big\}= \frac{\im}{2} g^{ij}\delta_b^a,&\nn \\ &\big\{   p_i, \psi{}^{ja}\big\} = \Gamma^j_{ik}\psi^{ka}, \; \; \big\{   p_i, \bpsi{}^{j}_a\big\} = \Gamma^j_{ik}\bpsi^{k}_a,\; \; \big\{   p_i, p_j\big\} = -2\im R_{ijkm}\psi{}^{ck}\,\bpsi{}^m_c.&
\eea
The Hamiltonian obtained is exactly the expected one:
\be\label{Hamcompl}
H = g^{ij}p_i p_j + g^{ij}W_i W_j + 4 \nabla_i W_j \, \psi{}^{ci}\bpsi{}^j_c - 4 \big[ \nabla_i F_{jkm} + R_{ijkm} \big]\psi{}^{ai}\bpsi{}_a^j \, \psi{}^{bk}\bpsi{}_b^m.
\ee
Therefore, to find the superfield version of the action of the system for the known curved WDVV solution $F_{ijk},W_m$ on the space with the metric $g_{ij}$, it is just required to solve the equation for the scalar $G$
\be\label{Geq}
\nabla_i \nabla_j G - F_{ijk}g^{km}\nabla_m G =g_{ij}.
\ee
Extended to the superfield, this $G$ would become a superfield Lagrangian that exactly reproduces the mechanical system given by the $g_{ij}$ and $F_{ijk}$ which, in turn, solve the curved WDVV equations.

For completeness, let us show that the supercharges that can be obtained using the superfield methods coincide with ones found in the Hamiltonian formalism \cite{cWDVV2}.  One may generate supersymmetry transformations of the components with the help of the formula
\be\label{susytr}
\delta_Q f = -\frac{1}{2} \big( \epsilon_a D^a + \bar\epsilon^a \bD_a   \big)f |_{\theta\rightarrow 0},
\ee
where after the calculation the equations of motion for the auxiliary fields and the definition of the momenta should be taken into account. These transformation laws for the components $x^i$ and $\psi^{ia}$ read
\bea\label{susytrcomp}
\delta_Q x^i &=& -\im \big(\epsilon_a \psi^{ia} + \bar\epsilon^a \bpsi_a^i   \big), \\
\delta_Q \psi^{ia} &=& \frac{\im}{2} \epsilon^a \big( \Gamma^{i}_{jk} + g^{im}F_{jkm}  \big)\psi^{jc}\psi_c^k - \frac{1}{2}\bar\epsilon{}^a g^{ij}p_j + \frac{\im}{2}\bar\epsilon{}^a g^{ij}W_j -\nn \\&&-\im\bar\epsilon{}^b \Gamma^i_{jk} \psi^{ja} \bpsi^k_b -\im \bar\epsilon_b g^{im}F_{jkm} \psi^{jb}\bpsi^{ak}. \nn
\eea
These transformations can also be reproduced via another formula involving the Dirac brackets
\be\label{susytrbrac}
\delta_Q f = \im \big\{ \epsilon_b Q^b + \bar\epsilon^b \bQ_b , f  \big\},
\ee
if the supercharges $Q^a$, $\bQ_b$ read
\be\label{Qfromtr}
Q^a = p_i \psi^{ia} + \im W_i \psi^{ia} + \im F_{ijk}\psi^{ci}\psi_c^j \bpsi^{ka}, \;\;\;
\bQ_a = p_i \bpsi^i_a - \im W_i \bpsi^i_a + \im F_{ijk} \bpsi_c^i \bpsi^{cj}\psi^k_a,
\ee
as expected.

Let us provide the superfield Lagrangian for the mechanics on the spheres as an example of the construction above. Let us take the curved WDVV solution \cite{cWDVV2} that exists on the spheres of any dimension,
\be\label{Fsphsol}
F_{ijk} = \frac{-4\rho^2 x^i x^j x^k}{\big( 1-\rho r^2 \big)\big(1+\rho r^2\big)^3}-\frac{2\rho \big( x^i \delta_{jk} +x^j \delta_{ik} + x^k \delta_{ij} \big)}{\big(1+\rho r^2\big)^3} + \frac{\partial_i \partial_j \partial_k \Big( \sum_{m=1}^N \big(x^m)^2 \log x^m   \Big)}{2\big( 1+ \rho r^2  \big)^2},
\ee
while the metric of the spheres reads
\be\label{sphmetric}
g_{ij} = \frac{\delta_{ij}}{\big( 1 + \rho r^2 \big)^2}, \;\; r^2 = \sum_i \big(x^i \big)^2, \; \; i=1\ldots N.
\ee
By solving equation \p{Geq} directly with $F_{ijk}$ \p{Fsphsol} and the metric \p{sphmetric} taken into account,
one can obtain, after replacing component $x^i$ with the superfield $\mx^i$,
\be\label{Gonsph}
G_{S^N} = \frac{\sum_i \big( \big(\mx^i\big)^2 \log \mx^i  \big)}{2\big( 1+\rho \mmr^2   \big)^2} - \frac{1}{4\rho}\mbox{ ArcTanh}\big( \rho \mmr^2  \big) - \frac{\mmr^2}{2}\frac{\log\big( 1 - \rho \mmr^2  \big)}{\big( 1 + \rho \mmr^2  \big)^2}.
\ee

Interestingly, one can obtain a similar expression not only by solving the differential equation \p{Geq} but by dimensional reduction from the sum of $N+1$ usual one-particle mechanics, as was proposed by Sergey Krivonos. Indeed, if $\mmu^A$ satisfy the conditions
\be\label{uconstr}
D^a D_a \mmu^A =0, \; \bD_a \bD^a \mmu^A =0, \; \big[ D^a, \bD_a \big] \mmu^A =0, \; A = 1 \ldots N+1,
\ee
then the sum of the actions of the conformal mechanics \cite{IKL} becomes the sum of usual mechanics after the substitution $u^A =\big( q^A \big)^2 $:
\be\label{confact}
S = -\frac{1}{16}\sum_A \int dt d^4 \theta\, \mmu^A \log \mmu^A = \frac{1}{16}\int dt \sum_A \big( {\dot q}{}^{A}{\dot q}{}^{A} + \mbox{fermions}  \big) .
\ee
Neglecting the fermions, one performs the reduction by enforcing $\sum_A \big(q^A\big)^2 = \rho^{-1}$. This constraint can be solved by
\be\label{stereogr}
 q^i = \frac{2 x^i}{1+ \rho r^2}, \; \; q^{N+1} = \frac{1}{\sqrt{\rho}}\frac{1-\rho r^2}{1+\rho r^2}.
\ee
Substituting this into \p{confact} without the fermions, one can obtain the bosonic action on the sphere $S^N$:
\be\label{sphbos}
S \approx \frac{1}{4}\int dt \frac{\sum_i \big(\dot x{}^i)^2}{\big( 1+ \rho r^2 \big)^2}.
\ee
It should be noted, however, that the reduction $\sum_A \big(\mq^A \big)^2 = \sum_A \mmu^A =\rho^{-1}$ is compatible with the constraints \p{uconstr} and therefore, preserves the supersymmetry. Substituting
\be\label{mmumx}
\mmu^i = \frac{4 \big(\mx^i \big)^2}{\big( 1+ \rho \mmr^2 \big)^2}, \;\; \mmu^{N+1} = \frac{1}{\rho}\frac{\big(1 -\rho \mmr^2  \big)^2 }{\big(1 + \rho \mmr^2  \big)^2 }
\ee
into the action \p{confact} and the constraints \p{uconstr}, one can obtain both the superfield Lagrangian \p{Gonsph} and the properly modified irreducibility conditions.

\section{Comments}
\begin{itemize}
\item
The real K\"{a}hler spaces were defined in \cite{Pash1} as spaces with the metric which is the second derivative of some prepotential,  seemingly by analogy with the well-known property of the usual complex K\"{a}hler spaces. However, in contrast to the complex case, this definition is not a covariant one and is difficult to use in the study of the properties of such spaces. Taking into account the results of the previous sections, it would be natural to use the curved WDVV equations to define such space. Indeed, {\it let us call the ``real K\"{a}hler space'' the Riemannian space equipped, together with the metric $g_{ij}$, with the third order Codazzi tensor $F_{ijk}$ that additionally satisfies the property}
\be\label{rKdef}
F_{ikp}g^{pq}F_{jlq} - F_{ilp}g^{pq}F_{jkq} + R_{ijkl} =0.
\ee
The symmetry property $F_{ijk}=F_{(ijk)}$ and the relation $\nabla_i F_{jkl} -\nabla_j F_{ikl}=0$ just define the third order Codazzi tensor.

One may actually show from this definition that $g_{ij}$ becomes $\partial_i\partial_j G$ in some coordinate system. Indeed, as was noted before, the connection $f^k_{ij} = \Gamma^k_{ij} + g^{km}F_{ijm}$ has zero Riemannian curvature \p{Feq2} as a consequence of the curved WDVV equations. Therefore, one may choose such a system of coordinates $x^\prime (x)$ that $f^k_{ij}=0$. In this system, $g^\prime_{k^\prime m^\prime}\Gamma^{\prime m^\prime}_{i^\prime j^\prime}=- F^\prime_{i^\prime j^\prime k^\prime}$, and therefore, also $\partial_{k^\prime} g^\prime_{i^\prime j^\prime}$ are symmetric in $i^\prime$, $j^\prime$, $k^\prime$ and $g^\prime_{i^\prime j^\prime}=\partial_{i^\prime} \partial_{j^\prime} G$.

Let us also note that the fact that it is possible to represent the curved WDVV equations as the zero curvature condition was already noted in \cite{cWDVV1}.

\item
Let us also note that  if $g_{ij}$ is part of some solution of the curved WDVV equations, the aforementioned result guarantees that the condition imposed on the superfield Lagrangian,
\be\label{Geq2}
\nabla_i \partial_j G - F_{ijk}g^{km}\partial_m G =g_{ij} \; \Leftrightarrow \partial_i \partial_j G - f^k_{ij}\partial_k G = g_{ij},
\ee
is always possible to satisfy. Indeed, it is enough to show that it possible to solve equations \p{Geq2} in a particular system of coordinates. But this condition becomes trivial to solve in the system of coordinates where $f^k_{ij}=0$, because in this system also $g_{ij} = \partial_i \partial_j {\widetilde G}$.

\item
It can be noted that for any given $f^k_{ij}$ and, consequently, any solution of the curved WDVV equations, it is possible to use not only a particular $G$ that satisfies equation \p{Geq}, but an {\it arbitrary} ${\widehat G}$. This, certainly, would produce a set of different systems. However, they could be described by the curved WDVV equations, too, as in the system of coordinates, where $f^k_{ij}=0$, they would become systems with the metric ${\hat g}_{ij} = \partial_i \partial_j {\widehat G}$, ${\widehat F}_{ijk} = \frac{1}{2} \partial_i \partial_j \partial_k {\widehat G} $, which are known to solve the curved WDVV equations. Though it is likely possible to use this idea to relate the solutions of the curved WDVV equations with each other, it does not generate a new class of solutions, and we do not consider it further in this paper.

\item There exists another type of manifolds, called the Frobenius manifolds, used by Dubrovin \cite{Dubr} to geometrize the WDVV equations. There exists some similarities between them: the Frobenius manifold is equipped with the metric and with the symmetric 3-tensor $c_{ABC}$ that is also the Codazzi tensor. There are a lot of dissimilarities, however. The metric of the Frobenius manifold is by definition flat, which makes the definition of the Codazzi tensor on it trivial.  Also, the Frobenius manifold is equipped with two vector fields. One is covariantly constant and is used to relate the metric to the one of the components of the $c_{ABC}$ equivalent to $g_{AB} = c_{1AB}$. The second, called the Euler vector field, is used to define the quasihomogeneity of the solutions of the equation. Both such fields are absent in the proposed definition of the real K\"{a}hler manifold. Additionally, the coefficients $c^A_{BC} = g^{AD}c_{DBC}$ are used to define the Frobenius algebras on the manifold, and the WDVV equation appears as an associativity condition on these algebras. At the same time, the real K\"{a}hler spaces are not associated with any algebras, and the equation  underlying them is nonhomogeneous.
\end{itemize}

\section*{Conclusion}
In this paper, we elaborated the simple method that allows to construct the superfield action for any supersymmetric mechanics given by a solution of the curved WDVV equations. This method involves a semitrivial modification of the multiplet constraints by the flat connection, which is composed of the Levi-Civita connection on the manifold and the third order Codazzi tensor. Also, to reproduce a given system, it is required to solve the differential equation, which the superfield Lagrangian satisfies. As a side effect of this result, it is proven that the metric of the manifold, on which the curved WDVV equations can be solved, can be represented as the second derivative of a prepotential in some distinguished system of coordinates. This, at the same time, provides a guarantee that the condition on the superfield Lagrangian can be solved, proves the complete equivalence of the Hamiltonian and the superfield methods of construction of systems of many linear multiplets, and allows us to propose the covariant definition of the real K\"{a}hler manifolds. Also, it was found that the curved WDVV equations naturally appear in the superfield formalism as conditions of consistency of the modified constraints.

\section*{Acknowledgments}
The work of N.K. was supported by the RSCF, grant 14-11-00598, and by the RFBR, grant 18-52-05002 Arm\_a. 
This paper is the further development of the joint works \cite{cWDVV1}, \cite{cWDVV2} together with Sergey Krivonos, Olaf Lechtenfeld, Armen Nersessian and Anton Sutulin. Author wishes to thank them for collaboration. 
Also, author wishes to acknowledge Sergey Krivonos for the method of obtaining the superfield actions on the spheres by dimensional reduction.


\section*{References}
\begin{thebibliography}{99}
\bibitem{cWDVV1}
Kozyrev N, Krivonos S, Lechtenfeld O, Nersessian A and Sutulin A 2017 Curved Witten-Dijkgraaf-Verlinde-Verlinde equation and {\cal N=4} mechanics {\it Phys.Rev}. {\bf D96} no.10, 101702 ({\it preprint arXiv:1710.00884})

\bibitem{cWDVV2}
Kozyrev N, Krivonos S, Lechtenfeld O, Nersessian A and Sutulin A 2018 {\cal N}=4 supersymmetric mechanics on curved spaces {\it Phys.Rev}. {\bf D97} no.8, 085015 ({\it preprint arXiv:1711.08734})

\bibitem{WittenWDVV}
Witten E 1990 On the Structure of the Topological Phase of Two-dimensional Gravity {\it Nucl.Phys.} {\bf B340} 281-332

\bibitem{DVVWDVV}
Dijkgraaf R, Verlinde H and Verlinde E 1991 Topological strings in $d < 1$ {\it Nucl.Phys.} {\bf B352} 59-86

\bibitem{Wyllard}
Wyllard N 2000 (Super)conformal many body quantum mechanics with extended supersymmetry {\it J.Math.Phys.} {\bf 41} 2826-2838 ({\it preprint arXiv: hep-th/9910160})

\bibitem{BGL}
Bellucci S, Galajinsky A and Latini E 2005 New insight into WDVV equation {\it Phys.Rev.} {\bf D71} 044023 ({\it preprint arXiv: hep-th/0411232})


\bibitem{Codazzi}
Liu H L, Simon U and Wang C P 1998 Higher order Codazzi tensors on conformally flat spaces {\it Contributions to Algebra and Geometry} {\bf 39} 329.

\bibitem{cWDVVSU21}
Kozyrev N, Krivonos S, Lechtenfeld O and Sutulin A 2018 $SU(2|1)$ supersymmetric mechanics on curved spaces {\it JHEP}. {\bf 1805} 175 ({\it preprint arXiv: 1712.09898})

\bibitem{Pash1}
Donets E, Pashnev A, Juan Rosales J and Tsulaia M 2000 N=4 supersymmetric multidimensional quantum mechanics, partial SUSY breaking and superconformal quantum mechanics {\it Phys.Rev.} {\bf D61} 043512 ({\it preprint arXiv:hep-th/9907224})

\bibitem{Pash2}
Donets E, Pashnev A, Rivelles V, Sorokin D and Tsulaia M 2000  N=4 superconformal mechanics and the potential structure of AdS spaces {\it Phys.Lett.} {\bf B484} 337-346 ({\it preprint arXiv:hep-th/0004019})

\bibitem{FILharms}
Fedoruk S, Ivanov E and Lechtenfeld O 2009 Supersymmetric Calogero models by gauging {\it Phys.Rev.} {\bf D79} 105015
({\it preprint arXiv:0812.4276})

\bibitem{KLD21alpha}
Krivonos S and Lechtenfeld O 2011 Many-particle mechanics with D(2,1:alpha) superconformal symmetry {\it JHEP} {\bf 1102} 042 ({\it preprint arXiv:1012.4639})

\bibitem{KLP}
Krivonos S, Lechtenfeld O and Polovnikov K 2009 $N=4$ superconformal n-particle mechanics via superspace {\it Nucl.Phys.}  {\bf B817} 265-283 ({\it preprint arXiv:0812.5062})

\bibitem{IKL}
Ivanov E, Krivonos S and Leviant V 1989 Geometric Superfield Approach To Superconformal Mechanics {\it J.Phys.} {\bf A22}  4201

\bibitem{Dubr}
Dubrovin B 1996 Geometry of 2D topological field theories In: {\it Springer LNM}, {\bf 1620}, 120--348



\end{thebibliography}
\end{document}